\documentclass[%
 aip,
 jmp,%
 amsmath,amssymb,
 preprint,%
prl]{revtex4-1}

\usepackage{graphicx}
\usepackage{dcolumn}
\usepackage{bm}
\usepackage[section]{placeins}
\begin{document}

\title[Non-Linear Langevin and Fractional Fokker–Planck Equations]
      {Non-Linear Langevin and Fractional Fokker–Planck Equations for Anomalous Diffusion by L\'{e}vy Stable~Processes} 

\author{Johan Anderson}%
\email{anderson.johan@gmail.com.}
\affiliation{ 
Department of Space, Earth and Environment, Chalmers University of Technology, 
SE-412 96 G\"{o}teborg, Sweden
}%

\author{Sara Moradi}%
\affiliation{ 
Laboratory for Plasma Physics---LPP-ERM/KMS, Royal Military Academy, 1000 Brussels, Belgium
}%

\author{Tariq Rafiq}
\affiliation{ 
Department of Mechanical Engineering and Mechanics, Lehigh University, Bethlehem, PA 18015, USA
}%

\date{\today}

\begin{abstract}
The~numerical solutions to a non-linear Fractional Fokker--Planck (FFP) equation are studied estimating the generalized diffusion coefficients. The~aim is to model anomalous diffusion using an FFP description with fractional velocity derivatives and Langevin dynamics where L\'{e}vy fluctuations are introduced to model the effect of non-local transport due to fractional diffusion in velocity space. Distribution functions are found using numerical means for varying degrees of fractionality of the stable L\'{e}vy distribution as solutions to the FFP equation. The~statistical properties of the distribution functions are assessed by a generalized normalized expectation measure and entropy and modified transport coefficient. The~transport coefficient significantly increases with decreasing fractality which is corroborated by analysis of experimental data.
\end{abstract}

\pacs{05.40Fb; 02.50Ey; 05.40-a}
\keywords{non-local theory; L\'{e}vy noise; Tsallis entropy; fractional Fokker--Plank equation; anomalous diffusion}
\maketitle


\section{Introduction}
In~magnetically confined (MC) plasma devices transport driven by turbulent fluctuations often severely limit the confinement time and thus impede the development of fusion as an alternative for electricity production. It is pertinent to understand and mitigate the effects of the turbulently driven transport where simplified models often are employed in order to elucidate the main features of the plasma turbulence. In~magnetised plasmas, it~is commonly accepted that turbulence is the primary cause of anomalous (i.e., elevated~compared to collisional) transport \mbox{\cite{horton, krommes2002}}\hspace{0pt}. It has been recognized that the nature of the anomalous transport processes is dominated by a significant ballistic or non-local component where a diffusive description is improper. The~turbulence in MC tokamak plasmas is anisotropic in the parallel and perpendicular length scales to the magnetic field and taps free energy from the pressure gradient that can drive fluctuations in electrostatic potential and \mbox{density~\cite{horton, krommes2002}}. The~super-diffusive properties are often ubiquitously found in plasmas, such~as the thermal and particle fluxes in the gradient region or in the Scrape-Off Layer (SOL) where the transport is dominated by the coherent structures (blobs)~\cite{carreras1996, carreras1999, milligen2005, sanchez2008, negrete2005, sanchez2006, hahm} and inherently possess a non-local character~\mbox{\cite{Zweben, Naulin, kaye, Lopez, Gentle, Mantica, van-Milligen}}. Moreover, there~is a large quantity of experimental evidence that density and potential fluctuations measured by Langmuir probes at different fusion devices support the idea that these fluctuations are distributed according to L\'{e}vy statistics. This was illustrated for example in~\cite{carreras1999}, where~probability density functions (PDFs) of the turbulence induced fluxes at the edge of the W7-AS stellarator were shown to exhibit power law characteristics in contrast to exponential decay expected from Gaussian statistics. Furthermore, the~experimental evidence of the wave-number spectrum characterized by power laws over a wide range of wave-numbers can be directly linked to the values of L\'{e}vy index $\alpha$ of the PDFs of the underlying turbulent processes. 
One widely used simplified model of a plasma is the Hasegawa--Wakatani model which was recently analysed by statistical methods in Reference~\cite{anderson2017}. It~was concluded that even simplified models may have components of fractionality stemming from the non-linear interactions and the generation of large scale modes such as zonal and shear flows. The~Hasegawa--Wakatani model allows for the electrons to dynamically and self-consistently determine the relationship between the density and the electrostatic potential through the turbulence. Moreover, fractal~features in transport have been observed in many experiments in many different fields of research. In~particular it has been found that there is strong evidence of non-local heat transport in JET plasmas~\cite{moradi2018}. In~this paper, fractal~features is synonymous to a system where power law statistics is found. Here it is important to keep in mind that, although~a simplified fractional transport model is used, it~indicates that there is a lack of physics in the current transport models based on the mean field theory, namely~the super-diffusive character of heat transport. 
Finding a proper kinetic description of dynamical systems with chaotic behaviour is one of the main problems in classical physics \mbox{
\cite{schlesinger1993, sokolov2002, klafter2005, metzler2000, metzler2004, mandelbrot1982, anderson2010, kim2009, moradipop2011, moradipop2012, anderson2014, moradipop2016}}\hspace{0pt}. Over the past two decades it has become obvious that behaviour much more complex  than standard diffusion can occur in dynamical Hamiltonian chaotic systems. In~principle, the~orbits in dynamical systems are always theoretically predictable since they arise as solutions to simple system of equations such as Newton's equations; however, these~orbits are sensitive to initial conditions and thus very small changes in initial conditions may yield widely different outcomes. From the macroscopic point of view, the~rapid mixing of orbits has been used as a motivation for assumptions of randomness of the motion and the random walk models \mbox{
\cite{schlesinger1993}}\hspace{0pt}. In~characterizing the diffusion processes in plasmas, the~starting point is often Brownian motion where the mean value vanishes, whereas~the second moment or variance grows linearly in time according to $\langle \delta x^2\rangle = 2 D t$. However, taking~into account the experimental data found in plasma experiments, it~is evident that many phenomena exhibit anomalous diffusion where variance grows non-linearly in time such that $\langle \delta x^2\rangle = 2 D t^{\alpha}$. The~reason an anomalous diffusion approach is needed is due to the restrictive assumptions of locality in space and time, and~lack of long-range correlations that is the basis of Brownian motion. There are two limits of interest for $\alpha$ where the first is super-diffusion with $\alpha > 1$ and the second is sub-diffusion with $\alpha < 1$. A super-diffusive description is most often appropriate for fusion plasmas. L\'{e}vy statistics describing fractal processes (L\'{e}vy index $\alpha$ where $0 < \alpha < 2$ ) lie at the heart of complex processes such as anomalous diffusion. L\'{e}vy statistics can be generated by random processes that are scale-invariant. This means that a trajectory will possess many scales, but~no single scale will be characteristic and dominate the process. Geometrically, this~implies the fractal property that a trajectory, viewed~at different resolutions, will~look self-similar. Such strange kinetics~\cite{schlesinger1993, mandelbrot1982} may be generated by accelerated or sticky motions along the trajectory of the random walk \mbox{
\cite{krommes2002}}\hspace{0pt}. Super-diffusivity may also occur as a result of variation in the step length of the motion, which~breaks the assumption that 
 a unique step length may, e.g., give~rise to long-range correlations in the dynamics generated by the presence of anomalously large particle displacements connecting otherwise physically disjoint domains. 

We note that, although~sub-diffusive processes are beyond the scope of the present work, its~properties have been studied in many different contexts where transport is often inhibited by sticky motion. Among sub-diffusive phenomena are  holes in amorphous semiconductors, where~a waiting time distribution with long tails has been introduced~\cite{montroll1973}. The~sub-diffusive processes within a single protein molecule have been described by generalized Langevin equation with fractional Gaussian noise~\cite{kou2004}. Turbulence and related anomalous diffusion phenomena have been observed in a wide variety of complex systems such as high energy plasmas, semiconductors, glassy~materials, nanopores, biological~cells, and~epidemic proliferation.

The~objective of the present paper is to explore the non-linear character of the fractional Fokker--Planck (FFP) equation resulting from a Langevin description driven by L\'{e}vy stochastic force with a non-linear interaction in the velocity. The~present work is based on previous efforts reported in Reference~\cite{anderson2014} and may provide new insights on the recent developments in the modelling  of the anomalous transport of charged particles in magnetised plasmas, such~as the non-local heat transport found in JET plasmas.

The~paper is organized in the following way: in Section~\ref{sec:FPL}, the~model is presented, and~the numerical results are shown and discussed in Section~\ref{sec:RES}. The~final section presents a discussion and~conclusions.

\section{The~Fokker--Planck and Langevin Equations} \label{sec:FPL}
Fractional kinetics is a powerful framework in describing anomalous transport processes exhibiting L\'{e}vy statistics. It is able to reproduce key aspects of anomalous transport including the non-Gaussian self-similar nature of the PDFs of particle displacement, and~the anomalous scaling of the moments. It has been shown that the chaotic dynamics can be described by using the FFP equation with coordinate fractional derivatives as a possible tool for the description of anomalous diffusion~\cite{zaslavsky}. Previous papers on plasma transport have used models including a fractional derivatives on phenomenological premises~\mbox{
\cite{sanchez2008,del-Castillo-Negrete2004,del-Castillo-Negrete2010}}\hspace{0pt}. Additionally, the~integro-differential nature of the fractional derivative operators allows the description of spatiotemporal nonlocal transport processes. In~particular, in~fractional diffusion, the~local Fourier--Fick's law is replaced by an integral operator in which the flux at a given point in space depends globally on the spatial distribution of the transported scalar and on the time history of the transport process. Using fractional generalizations of the Liouville equation, kinetic~descriptions have been developed~\mbox{
\cite{zaslavsky2002, tarasov2005, tarasov2006}}\hspace{0pt}. The~currently applied model is based on the Langevin equation with a L\'{e}vy-stable noise term, where~the applied noise exhibits a power law tail~\mbox{\cite{levy, seshadri}}.
The~generalized Central Limit Theorem for L\'{e}vy-stable processes is a particular weak-convergence theorem in probability theory. It expresses the fact that a sum of many independent and identically distributed (i.i.d.) random elements, or~alternatively, random~elements with specific types of dependence, will~tend to be distributed according to one of a small set of attractor distributions. There are here two cases of special interest: the first is when the variance of the i.i.d. variables is finite and the attractor distribution is then a normal distribution, and~the second is 
where the sum of a number of i.i.d. random elements with power law tail distributions decreasing as $|x|^{-\alpha-1}$ where $0<\alpha<2$ (therefore having infinite variance) 
will tend to a L\'{e}vy-stable distribution with a fractality index of $\alpha$ as the number of elements in the set increases.

The~motion of a colloidal particle can be described by the Langevin equation in the case of Brownian motion and it will take the form
\begin{equation}
\frac{d}{dt} v = -\nu v + A(t), \label{eq:1.1}
\end{equation}
where $v$ is the speed of the particle, $- \nu v$ is the friction, and~$A(t)$ is the white stochastic force such that $\langle A(t) A(t^{\prime}) \rangle = 2 D \delta(t - t^{\prime})$. Moreover, by~assuming that $A(t)$ is a Gaussian stochastic force, a~Maxwellian velocity distribution may be obtained and would lead to the standard Fokker--Planck (FP) equation describing the evolution of the distribution function:
\begin{equation}
\frac{\partial}{\partial t} P + v \frac{\partial P}{\partial r} + \frac{F}{m} \frac{\partial P}{\partial v} =  \nu \frac{\partial}{\partial v}(v P) + D \frac{\partial^2 P}{\partial^2 v}. \label{eq:1.2}
\end{equation}

Here $P$ is the distribution function, $v$ is the velocity, $F$ is an external force, e.g., the~electromagnetic force, $m$ is the mass, $\nu$ is the friction, and~$D$ is the diffusion coefficient. The~corresponding reduced FP equation , where~the Lorentz force is neglected, to~the Langevin equation is
\begin{equation}
\frac{\partial}{\partial t} P =  \nu \frac{\partial}{\partial v}(v P) + D \frac{\partial^2 P}{\partial^2 v}, \label{eq:1.3}
\end{equation}
which yields to the stationary state Maxwellian velocity distribution for $P(v)$ \mbox{ {\cite{fogedby1994, fogedby19942}}}.   
However, if~$A(t)$ is a stochastic noise with the properties of a L\'{e}vy-stable process, the~FP equation has to be modified in order to accommodate for power law tails of the form $P(v) \propto v^{-\alpha - 1}$ for a L\'{e}vy stable with fractional index $\alpha$. The~FFP equation becomes
\begin{equation}
\frac{\partial}{\partial t} P(v,t) =  \nu \frac{\partial}{\partial v}(v P(v,y)) + D \frac{\partial^{\alpha} P(v,t)}{\partial^{\alpha} |v|}  \label{eq:1.4}
\end{equation}
where $0< \alpha \leq 2$ and $|v|< \infty$. The~time-dependent solution is readily found in the Fourier space where the fractional Riesz operator in 1 + 1D can be transformed to
\begin{equation}
\frac{\partial}{\partial t} \hat{P}(k,t) =  - \nu k \frac{\partial}{\partial k}(\hat{P}(k,t)) - D |k|^{\alpha} \hat{P}(k,t) \label{eq:1.5}
\end{equation}
where the Fourier transformed distribution function can be determined to be
\begin{equation}
\hat{P}(k,t) =  \exp(-\frac{D |k|^{\alpha}}{\nu \alpha}(1 - \exp(-\nu \alpha t))). \label{eq:1.6}
\end{equation}

The~fractional Riesz derivative is defined through its Fourier transform $_{-\infty}\hat D_x^{\mu} f(x) = \frac{\partial^{\mu} f(x)}{\partial^{\mu} |x|} = - |k|^{\mu} f(k)$, see, e.g.,~\cite{metzler2000} for more information. Here it should be noted that the time derivative only introduces a relaxation time dependent on the friction and the fractionality $\alpha$, where~a smaller $\alpha$ yield a longer relaxation time.    

\begin{figure}[ht]
\centering
\includegraphics[width=7 cm, height=5.5cm]{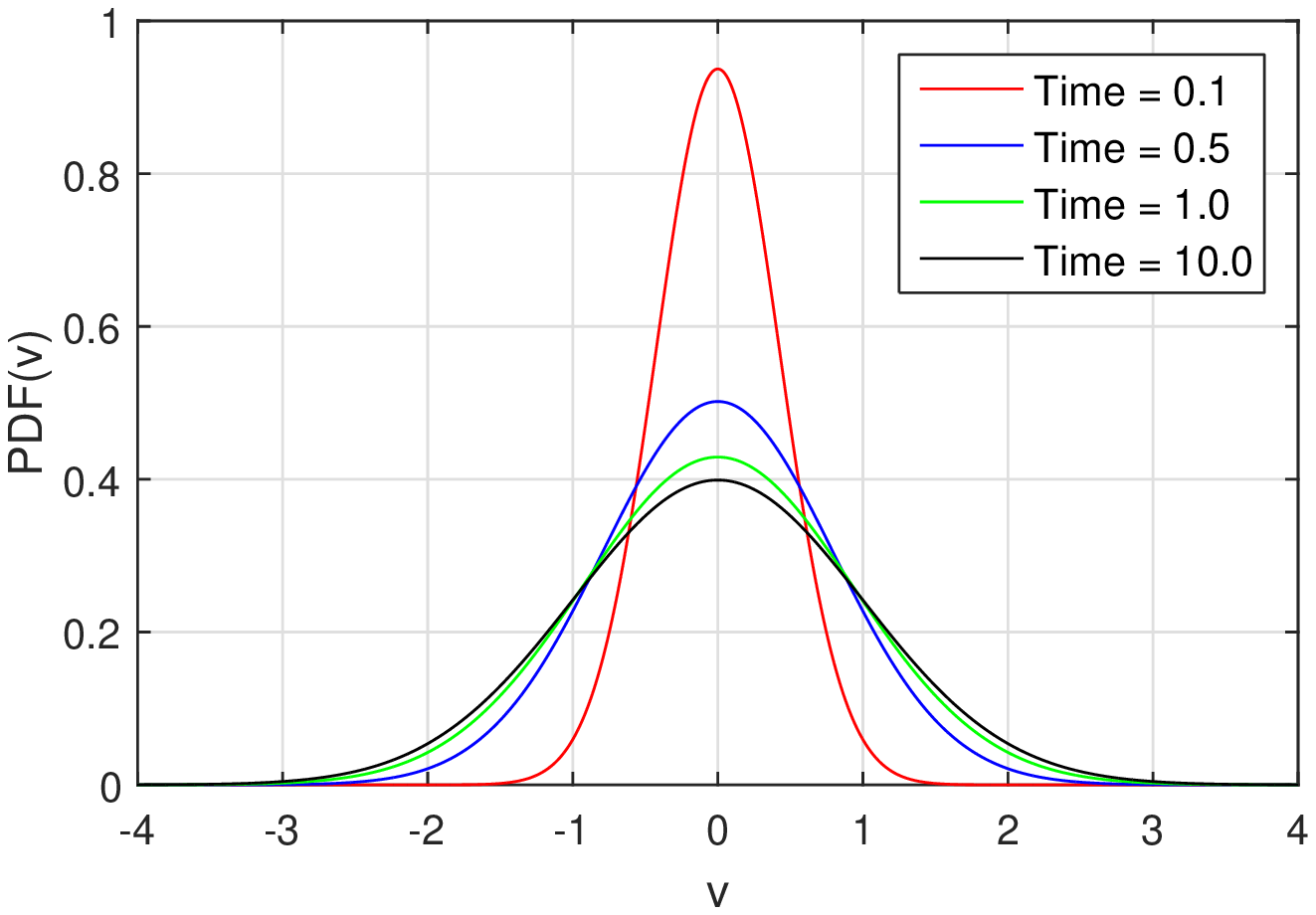}
\includegraphics[width=7 cm, height=5.5cm]{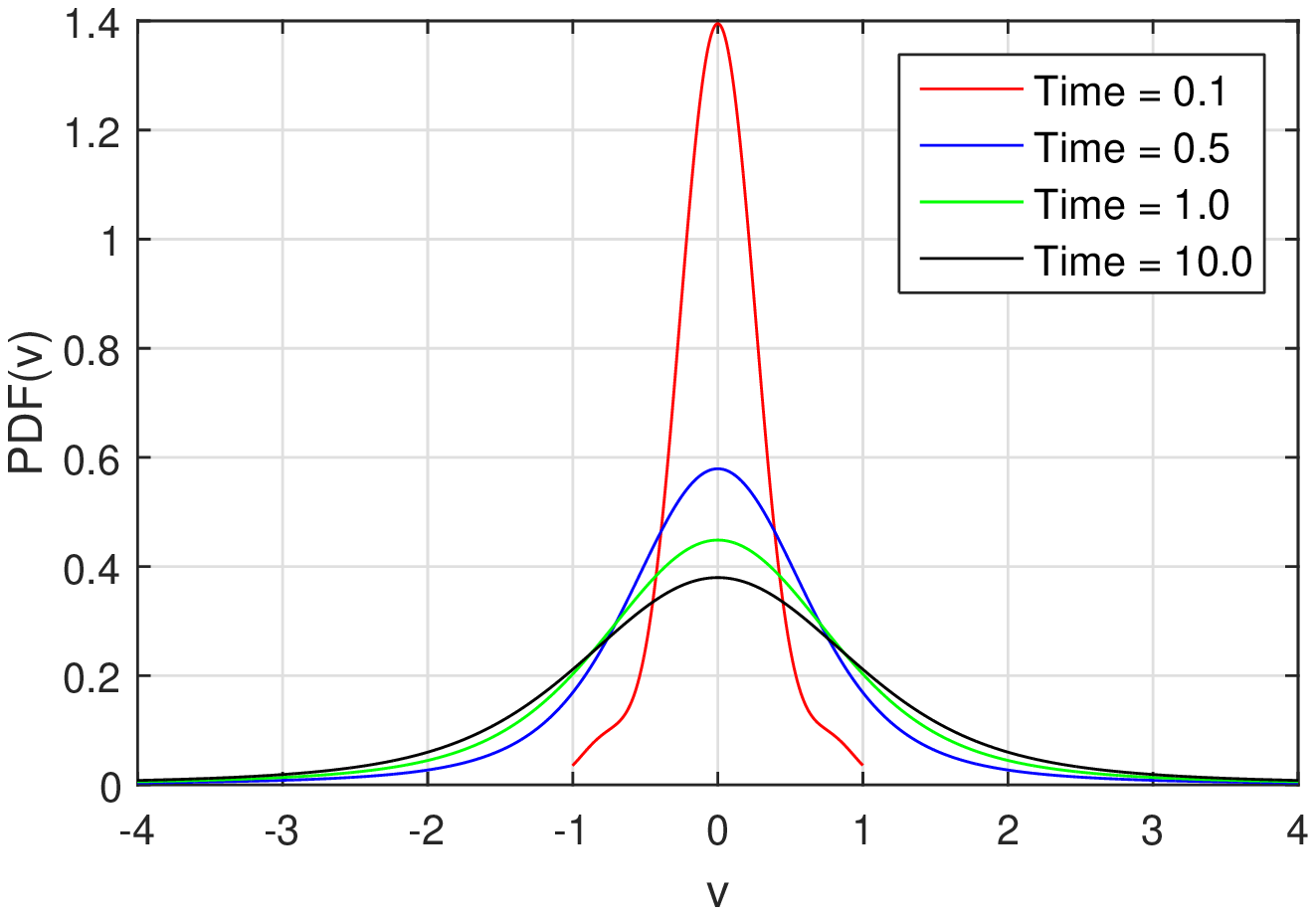}
\caption{The~probability density function  (PDF)  of velocity computed by the inverse Fourier transform of Equation~\eqref{eq:1.6} with $\alpha = 2.0$ (\textbf{left}) and $\alpha = 1.5$ (\textbf{right}) for $t=0.1, 0.5, 1.0, 10.0$.} \label{fig:1}
\end{figure}

In~Figure~\ref{fig:1}, the~exponentially fast relaxation of the velocity PDFs with time is displayed. The~PDFs of a Gaussian ($\alpha = 2.0$) and for a PDF with fractional index $\alpha = 1.5$ for times $t = 0.1, 0.5, 1.0$, and~$10.0$ are computed numerically. We note that, at~$t=10.0$, the~PDFs are close to the stationary state PDF, whereas~the time evolution of the PDF depends on the fractional index $\alpha$ such that the relaxation process is slower for a PDF with a lower fractional index. In~general, the~distributions found for the $\alpha = 1.5$ have more pronounced tails and sharper peaks, whereas, in~the $\alpha = 2.0$ case, the~system has a shorter relaxation time.
  
\section{Results} \label{sec:RES}
The~aim of the present paper is to look into the effects of a non-linear interaction in the Langevin equation, but~it is here assumed that we can neglect the time dependence , i.e., the stationary state PDF ($\frac{dP}{dt} = 0$), and~the FFP Equation can be written as
\begin{equation}
0 = \nu \frac{\partial}{\partial v} \left((v + \beta v^3) P \right) + D \frac{\partial^{\alpha}}{\partial^{\alpha} |v|} P. \label{eq:1.7}
\end{equation}

Here $\nu$, $\beta$, and~$D$ are constants. The~equation is obtained by inclusion of the quartic potential, leading~to the addition of a term of third order of the form $\beta v^3$. The~main effect of retaining the temporal dynamics is to introduce a relaxation time. In~the current model, square~and quartic terms will be retained. The~properties of the current non-linear terms are analogous to a potential well with square and quartic terms. Note that even terms in the potential provide proper stable equilibria, whereas~odd terms yield an unstable equilibrium; thus, the~third and fifth order terms are neglected. The~Equation~(\ref{eq:1.7}) is directly integrated by using a predictor---corrector method according to Adams-Bashforth-Moulton~\cite{diethelm}.

To find an analytical solution of the original Langevin equation, the~Fourier transform can be~used; 
\begin{equation}
\nu \left[ \frac{\partial}{\partial k}  - \beta \frac{\partial^3}{\partial k^3}\right] \tilde{P} + D |k|^{\alpha - 1} \tilde{P} = 0.
\end{equation} \label{eq:1.7f}

The~found equation is a third order ordinary differential equation with variable coefficients. The~general solutions to Equation~(\ref{eq:1.7f}) can only be determined by numerical means however a similar system was investigated in Reference~\cite{kim2009} suggesting a PDF proportional to $exp(-a*v^4)$, where~$a$ is a constant.  Furthermore, it~is also possible to find an analytical solution for the tail of the PDF to leading order by using the Wentzel–Kramers–Brillouin (WKB) approximation for small values of $\beta$. The~WKB anzats is to assume a series solution to the Fourier transformed equation (\ref{eq:1.7f}), of~the form $\tilde{P}(k) = exp{(\frac{1}{\epsilon} \sum_{n=0}^{\infty} \epsilon^n S_n(k))}$, here~$\epsilon$ will be taken small and to be determined in terms of $\beta$. It is then found that, the~leading order tail contribution corroborates the findings in~\cite{kim2009} for $\alpha = 2.0$. We note that the real space distribution function is convergent for $\beta > 0$ and can only in general be obtained by numerical integration, and~is here solved by using method described above. We note that there are three different interesting regimes: the first is where the diffusion is much larger than the quartic potential strength $D/\nu >> \beta v^2$, the~second is where the diffusion is comparable to the quartic potrential strength $D/\nu \sim \beta v^2$, and~the third is where the diffusion is negligible to the quartic potential strength $D/\nu << \beta v^2$. In~the third regime, the~PDFs become may be expected to have similarities to the results found in~\cite{kim2009} for $\alpha = 2.0$ where $P(v) \sim \exp (- a v^4)$ for some constant $a$. The~values used in this study are chosen to illustrate these three regimes of interest. Note that the non-linear interaction, i.e., the $\beta v^3$ term introduces three different regimes with richer dynamics which is in contrast to what was found in Reference~\mbox{\cite{anderson2014}}\hspace{0pt}. In~any linear fractal model based on the L\'{e}vy statistics the power law tails of the velocity PDF will be $P(v) \propto |v|^{-\alpha-1}$. Even more interestingly, in~non-linear models the precise scaling of the PDF tails are still open.

\begin{figure}[ht]
\centering
\includegraphics[width=13 cm]{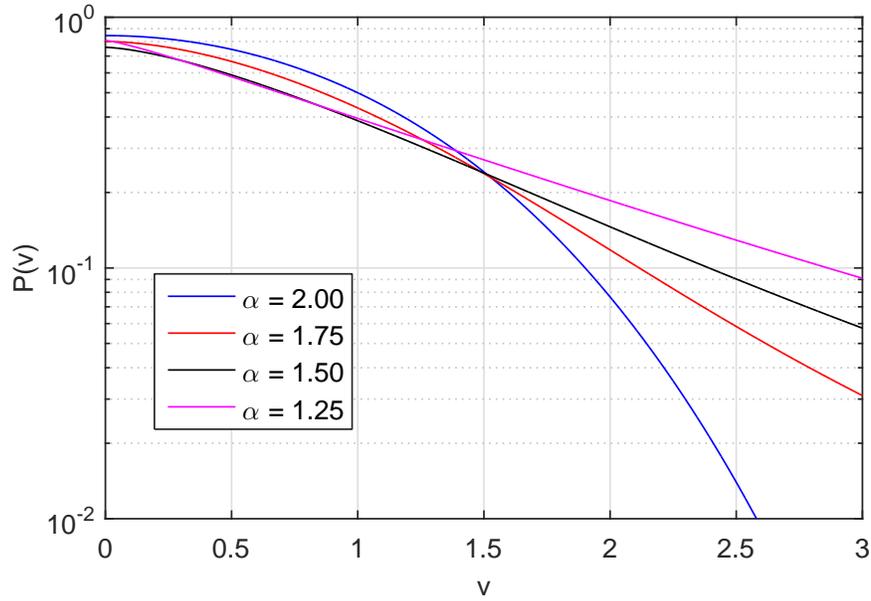}
\caption{The~PDF of velocity computed by integration of Equation~(\ref{eq:1.7}) with with $\alpha = 1.25$ (magenta line), $\alpha = 1.5$ (black line), $\alpha = 1.75$ (red line), and~$\alpha = 2.0$ (blue line) for $D/\nu = 1.0$ and $\beta = 0.1$.} 
\label{fig:2}
\end{figure}

{In~Figures~\ref{fig:2}--\ref{fig:4}, the~numerically found PDFs, by~solving Equation~(\ref{eq:1.7}) by the Adams-Bashforth-Moulton method~\cite{diethelm}, are~shown for the three different regimes: $D/\nu = 1.0$ and $\beta = 0.1$, $D/\nu = 1.0$ and $\beta = 1.0$, and~ $D/\nu = 0.1$ and $\beta = 1.0$, respectively. The~resolution in $v$ is $2^{-12}$ except in the case of $\alpha = 1.25$ for $D/\nu = 1.0$ and $\beta = 1.0$ where the resolution is increased to $2^{-20}$, however~increased resolution would not change the $P(v)$ in any significant way for smaller |v|.
As expected, in~Figure~\ref{fig:2} almost independently of $\alpha$ where $D/\nu > \beta v^2$, the~PDFs exhibit power law tails, although~in the case of $\alpha = 2$ some exponential behaviour is observed at the tail of the distribution (for large modulus of {velocity}~$|{v}|$). In~Figures~\ref{fig:3} and~\ref{fig:4}, we~find more pronounced tails in particular in the low-$\alpha$ case. In~the low-$\alpha$ case, the~fractal term dominates the dynamics. We note that the PDFs displayed in Figure~\ref{fig:3} exhibit a hybrid between fractal and Gaussian behaviour ($D/\nu \sim \beta v^2$), where~in some cases the PDF is retains some Gaussian behaviour, which~is  particularly visible for small $|v|$. In~the regime $D/\nu << \beta v^2$ the non-Gaussian effects of the PDFs are clearly visible in Figure~\ref{fig:4}. The~PDFs are used to evaluate the dynamics of the system in terms of Tsallis' statistical mechanics where $q$-entropy and $q$-energy determines the properties of the system for the three different regimes: $D/\nu = 1.0$ and $\beta = 0.1$, $D/\nu = 1.0$ and $\beta = 1.0$, and~ $D/\nu = 0.1$ and $\beta = 1.0$ in Figures~\ref{fig:5} and~\ref{fig:6}, respectively. The~$q$, $q$-entropy, and~$q$-energy values are determined by the following relations (see References \mbox{
\cite{tsallis1995lnp, tsallis1996, tsallis1998, hamza, barkai, angulo, balasis2011, pavlos2012, pavlos20122, tsallis1995, prato1999, anderson2014}}\hspace{0pt}):}
\begin{eqnarray}
q & = & \frac{3 + \alpha}{1 + \alpha} \label{eq:1.10} \\
S_q & = & \frac{1-\int dv (F(v))^q}{q-1} \label{eq:1.11}\\
\langle v^2 \rangle_q & = & \frac{\int dv (F(v))^q v^2}{\int dv (F(v))^q}. \label{eq:1.12}
\end{eqnarray}

{The~entropy and $q$-entropy are displayed in Figure~\ref{fig:5}. A maximum in the entropy, computed~according to $S_B = \int dv P(v) lnP(v)$, is~found, whereas~almost constant $q$-entropy, computed~according to Equation~(\ref{eq:1.11}), is~found with increasing $q$, where~$q$ is determined according to Equation~(\ref{eq:1.10}). An increasing trend with importance of non-local effects are also visible where the entropy and $q$-entropy is increased in the cases with $\beta = 1.0$, which~is in the regime where the fractal nature is more prominent. In~Figure~\ref{fig:6}, a~decreased energy and increased $q$-energy with increasing $q$ for $\beta = 1.0$ and small $D/\nu$ case is found.}

\vspace{-6pt}

The~interpretation of this strange kinetics has to be based on the results from experimental data since there is no first principle method to compute the value of $\alpha$ and thus $q$ is indeterminable. However, recent~findings suggest that JET plasmas have a significant degree of super-diffusive transport with an $\alpha < 2$, and~it was found that this super-diffusive transport is slightly different for the ion and electron channels~\cite{moradi2018}. The~analysis is based on a power balance where a large set of JET shots are used whereby a distribution in $\alpha$ can be obtained with a mean value of approximately $1$, suggesting~that a convective model would be more appropriate with $q \approx 2$. The~diffusion coefficient can be estimated by the velocity autocorrelation functions according to the Kubo formula, but~such an estimate looks at the ratio of the generalized diffusion coefficient ($D_{\alpha}$) and the Brownian diffusion coefficient $D_0 = D(\alpha = 2.0)$, using~the tempered $q$-velocity correlators{, computed~by Equation~(\ref{eq:1.12}).

\vspace{-12pt}

\begin{figure}[ht]
\centering
\includegraphics[width=13 cm]{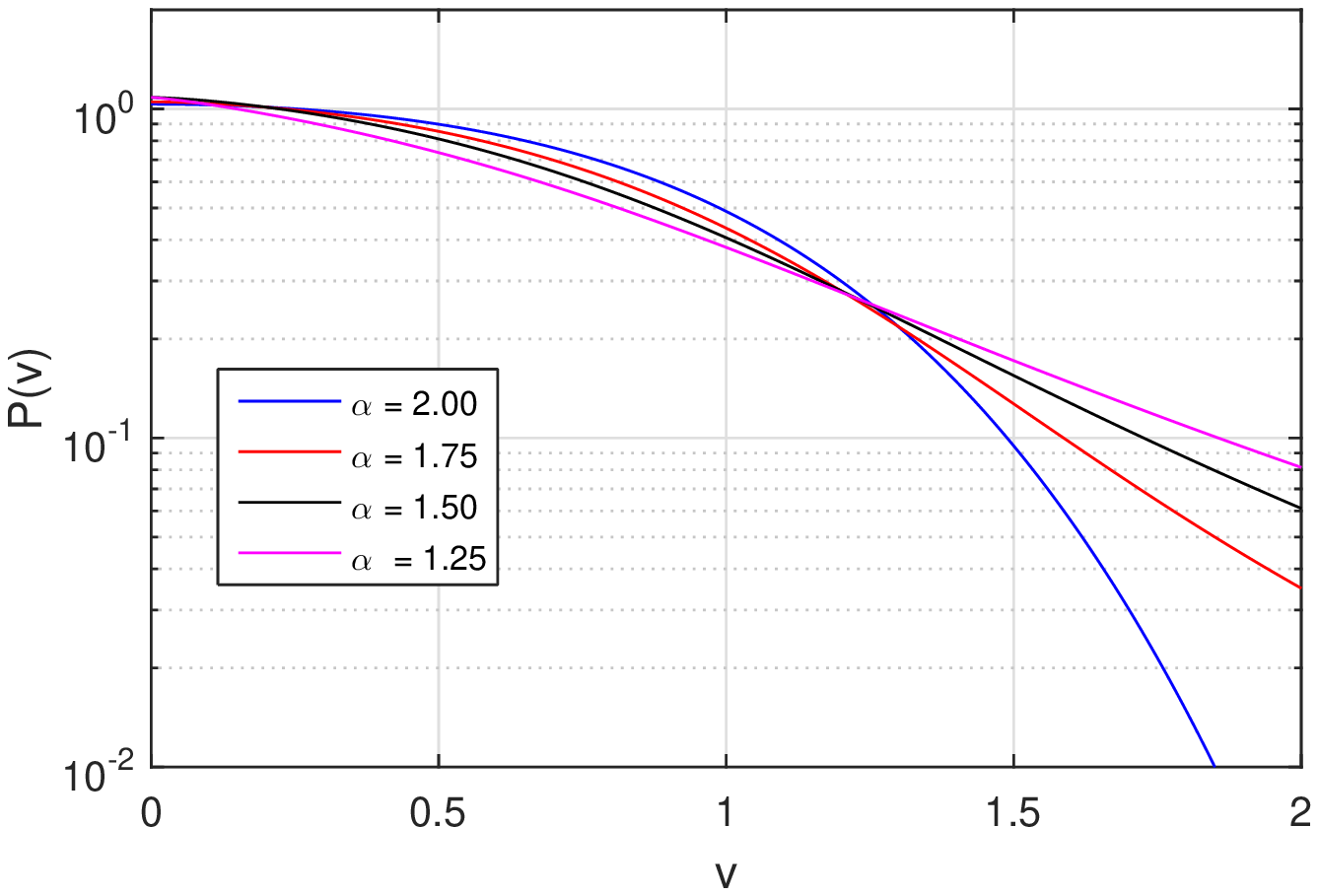}
\caption{The~PDF of velocity computed by integration of Equation~(\ref{eq:1.7}) with $\alpha = 1.25$ (magenta line), $\alpha = 1.5$ (black line), $\alpha = 1.75$ (red line), and~$\alpha = 2.0$ (blue line) for $D/\nu = 1.0$ and $\beta = 1.0$.} \label{fig:3}
\end{figure}

\vspace{-14pt}

\begin{figure}[ht]
\centering
\includegraphics[width=13 cm]{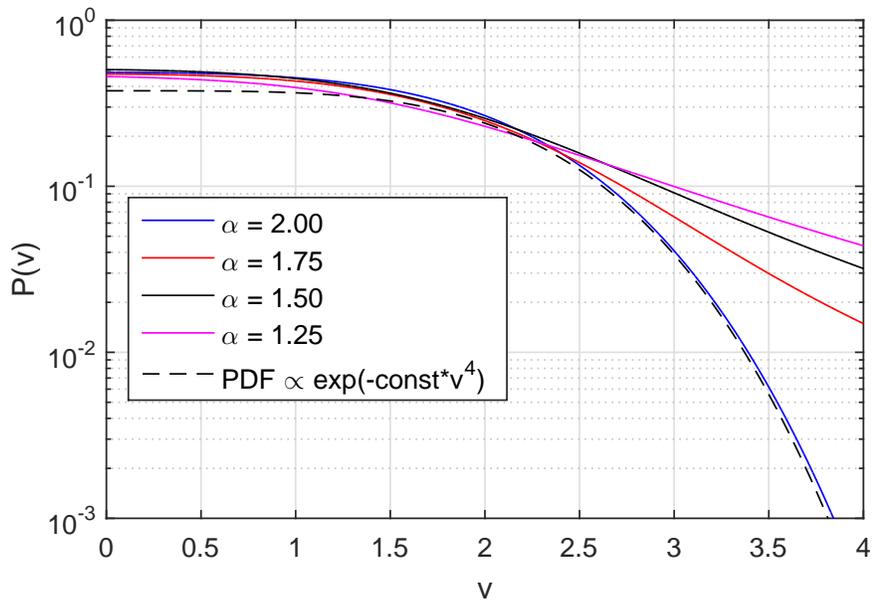}
\caption{The~PDF of velocity computed by integration of Equation~(\ref{eq:1.7}) with with $\alpha = 1.25$ (magenta line), $\alpha = 1.5$ (red line), $\alpha = 1.75$ (red line), and~$\alpha = 2.0$ (blue line) for $D/\nu = 0.1$ and $\beta = 1.0$.} \label{fig:4}
\end{figure}

We find that the ratio of the diffusion coefficients increases with smaller $\alpha$ and significantly increases in the regime where fractality is pronounced, as~shown in Figure~\ref{fig:7}. Interestingly enough, in~the analysis presented in ~\cite{moradi2018} it is evident that in the cases with increased transport a lower value of $\alpha$ is obtained, indicating~a strong non-diffusive component or equivalently, a~significantly increased transport where processes following L\'{e}vy statistics dominate the transport. The~qualitative increase in the generalized transport coefficient $D_q$ is thus qualitatively corroborated by what is seen experimentally using the power balance analysis.

\vspace{-6pt}

\begin{figure}[ht]
\centering
\includegraphics[width=13cm]{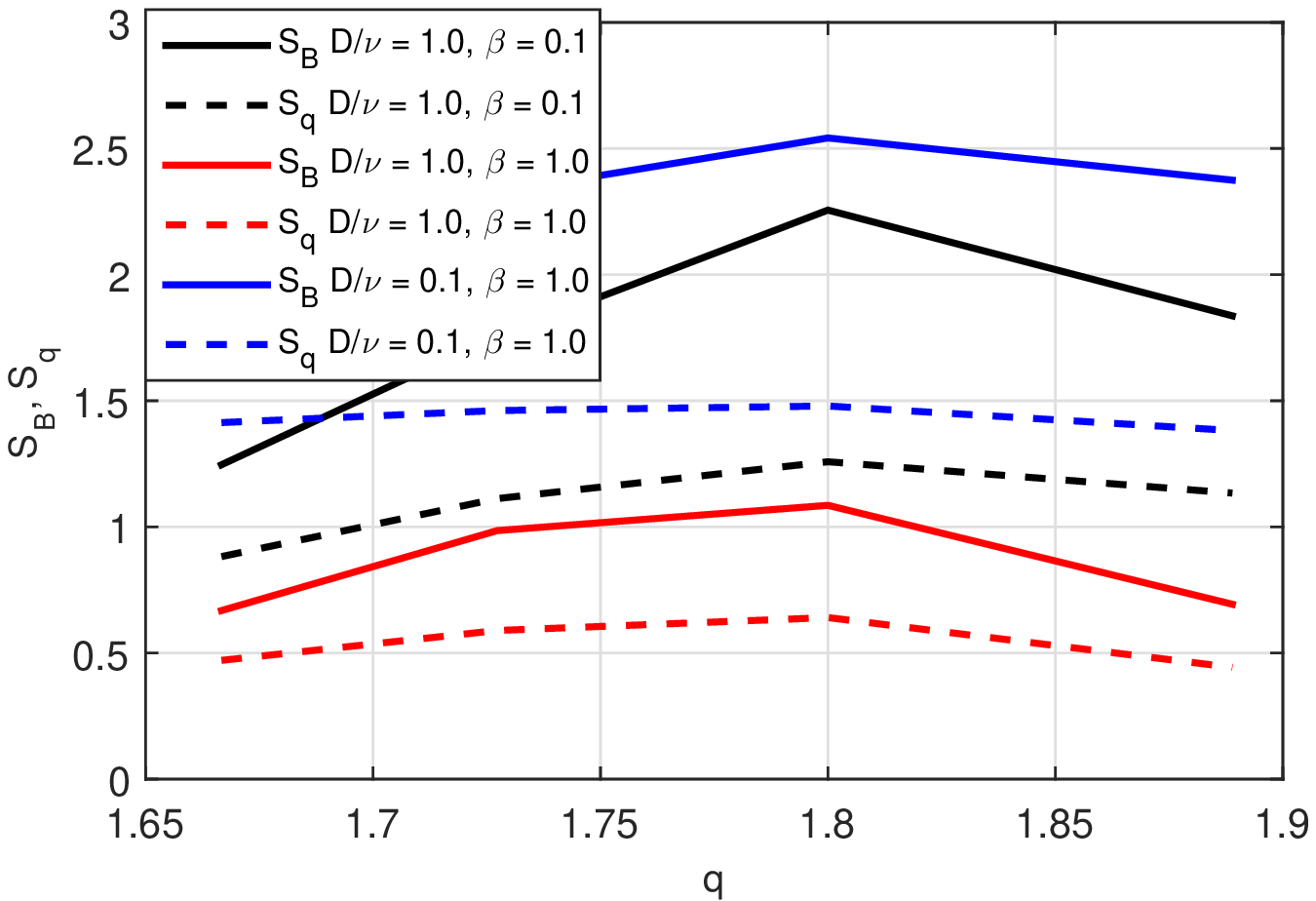}
\caption{The~Boltzmann--Gibbs entropy and the Tsallis' entropy as functions of the fractality index $q$ for $D/\nu = 1.0$ and $\beta = 0.1$ (solid black line and dashed black line, respectively), $D/\nu = 1.0$ and $\beta = 1.0$ (solid red line and dashed red, respectively), $D/\nu= 0.1$ and $\beta = 1.0$ (solid blue line and dashed blue line, respectively).} \label{fig:5}
\end{figure}

\vspace{-18pt}

\begin{figure}[ht]
\centering
\includegraphics[width=13cm]{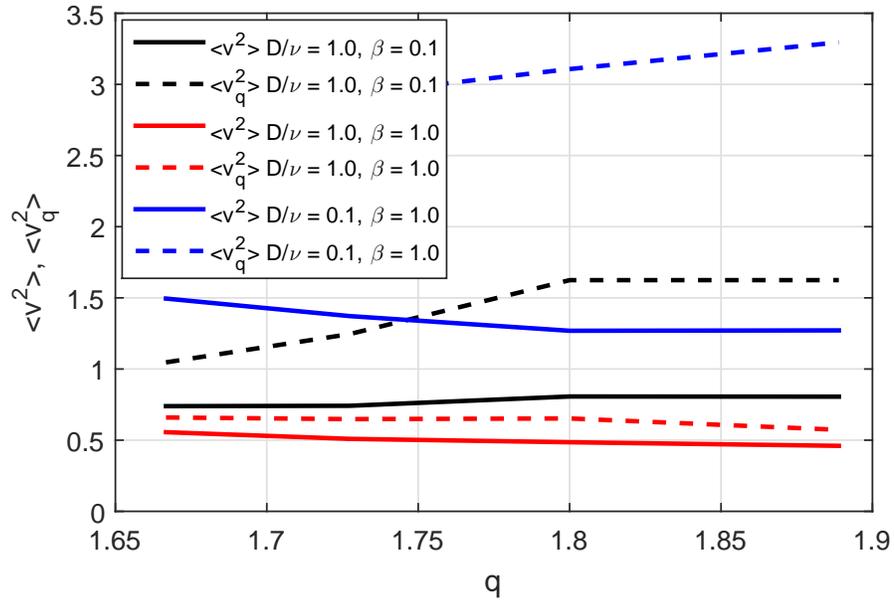}
\caption{The~energy and the generalized $q$-energy as functions of the fractality index $q$ for $D/\nu = 1.0$ and $\beta = 0.1$ (solid black line and dashed black line, respectively), $D/\nu = 1.0$ and $\beta = 1.0$ (solid red line and dashed red, respectively), $D/\nu= 0.1$ and $\beta = 1.0$ (solid blue line and dashed blue line,~respectively).} \label{fig:6}
\end{figure}

\vspace{-6pt}

\begin{figure}[ht]
\centering
\includegraphics[width=13cm]{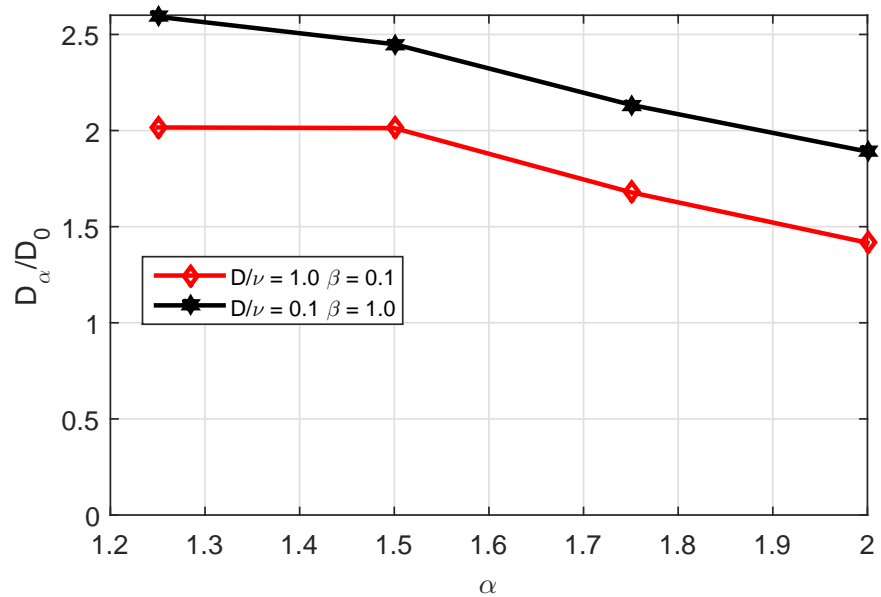}
\caption{The~ratio of the generalized diffusion coefficient ($D_{\alpha}$) and the Brownian diffusion coefficient as functions of the fractality index $\alpha$ for $D/\nu = 1.0$ and $\beta = 0.1$ (red line) and $D/\nu = 0.1$ and $\beta = 1.0$ (black line).} \label{fig:7}
\end{figure}

\section{Discussion and Conclusions}
Understanding anomalous transport in MC plasmas is an outstanding issue in controlled fusion research. It is commonly accepted that, in~these plasmas, turbulence~is the primary cause of anomalous (i.e., elevated~compared to collisional) transport. It has also been recognized that the nature of the anomalous transport processes is dominated by a significant ballistic or non-local component where a diffusive description is improper. A satisfactorily understanding of the non-local features as well as the non-Gaussian PDFs found in experimental measurements of particle and heat fluxes is still lacking \mbox{
\cite{anderson2010, kim2009}}\hspace{0pt}, but~there has been some recent progress in this direction. Fractional kinetics has been put forward for building a more physically relevant kinetic description for such dynamics. In~these situations, kinetic~descriptions, which~arise as a consequence of averaging over the well-known Gaussian and Poissonian statistics (for diffusion in space and temporal measures, respectively), seem~to fall short in describing the apparent randomness of dynamical chaotic systems~\cite{schlesinger1993}. This is due to the restrictive assumptions of locality in space and time, and~the lack of long-range correlations that is the basis of these descriptions.

In~magnetised plasma experiments, a~predator--prey system exists with avalanches (strong driver of transport) and zonal flows (sheared flows that decorrelate turbulent eddies reducing transport). It has been suggested that an Fractional Fokker--Planck Equation on a comb-like potential background can be applied where meso-scale transport events (avalanching) occurs in between regions of strong zonal flow activity (see Milovanov and Rasmussen~\cite{milovanov}). This method is straightforward for applications in this setting; by assuming the used potential in between the zonal flow regions, it~is suggested that the potential should be of degree 4 (or higher), as~has been used here.

Although there has previously been some criticism on the appropriateness of using the Tsallis method in describing processes with L\'{e}vy statistics, this~is mainly concerning descriptions based on fractality in coordinate space not in velocity space. However, the~aim of the present work was to shed light on the non-extensive properties of the velocity space statistics and characterization of the fractal processes by estimating the generalized diffusion coefficients of the FFP equation in terms of Tsallis statistics. Jespersen et al.~\cite{jespersen} showed an example of the Langevin equation with a harmonic potential, and~the Tsallis $q$-statistics had limited usefulness. The~reason for this is that, using~the variational calculus of Equation (\ref{eq:1.11}) with the appropriate constraints, the~relation between $\alpha$ and $q$ is $\alpha = \frac{4-2q}{q-1}$, which~is different from Equation (\ref{eq:1.10}) and thus cannot reproduce the correct scaling. They then concluded that the Tsallis entropy was not the appropriate framework for L\'{e}vy flights in a harmonic potential described by the generalized FP equation. However, this~limitation seems not to impede the usefulness of the application of Tsallis entropy on this Langevin equation where the correct scaling is obtained.

In~summary, we~have employed an FFP equation to find the PDFs and studied the $q$-entropy and $q$-energies in this system with a non-linear interaction in the FFP equation. We found a significantly elevated diffusion coefficient, which~is qualitatively similar to what was expected in light of the analysis of the experimental data.

\renewcommand\bibname{References}


\begin{thebibliography}{999}

\bibitem{horton} {Horton, W. {\em Turbulent Transport in Magnetized Plasmas}; { World Scientific: Danvers, MA, USA, 2017}; {ISBN 978-981-3225-88-6.}}
\bibitem{krommes2002} {Krommes, J.A.} {Fundamental statistical descriptions of plasma turbulence in magnetic fields.} {\em Phys. Rep.} {\bf 2002}, {\em 360}, 1--352.
\bibitem{carreras1996} {Carreras, B.A.; Hidalgo, C.; Sanchez, E.; Pedrosa, M.A.; Balbin, R.; Garcia, C.I.; van Milligen, B.; Newman,~D.E.; Lynch, V.E.} {Fluctuation‐induced flux at the plasma edge in toroidal devices.} {\em Phys. Plasmas} {\bf 1996}, {\em 3}, 2664--2672.
\bibitem{carreras1999} Carreras, B.A.; van Milligen, B.; Hidalgo, C.; Balbin, R.; Sanchez, E.; Cortes, I.G.; Pedrosa, M.A.; Bleuel, J.; Endler, M. Self-similarity properties of the probability distribution function of turbulence-induced particle fluxes at the plasma edge. {\em Phys. Rev. Lett.} {\bf 1999}, {\em 83}, 3653--3656.  
\bibitem{milligen2005} {Van Milligen, B.P.; Sanchez, R.; Carreras, B.A.; Lynch, V.E.; LaBombard, B.; Pedrosa, M.A.; Hidalgo, C.; Gonçalves, B.; Balbín, R.; The~W7-AS Team.} {Additional evidence for the universality of the probability distribution of turbulent fluctuations and fluxes in the scrape-off layer region of fusion plasmas.} {\em Phys.~Plasmas} {\bf 2005}, {\em 12}, 52501--52507.
\bibitem{sanchez2008} {Sanchez, R.; Newman, D.E.; Leboeuf, J.N.; Decyk, V.K.; Carreras, B.A.} {Nature of Transport across Sheared Zonal Flows in Electrostatic Ion-Temperature-Gradient Gyrokinetic Plasma Turbulence.} {\em Phys. Rev. Lett.} {\bf 2008}, {\em 101}, 205002--205004.
\bibitem{negrete2005} {del-Castillo-Negrete, D.; Carreras B.A.; Lynch, V.E.} {Front Dynamics in Reaction-Diffusion Systems with Levy Flights: A Fractional Diffusion Approach.} {\em Phys. Rev. Lett.} {\bf 2005}, {\em 94}, 18302--18304.
\bibitem{sanchez2006} {Sanchez, R.; Carreras, B.A.; Newman, D.E.; Lynch, V.E.; van Milligen, B.P.} {Renormalization of tracer turbulence leading to fractional differential equations.} {\em Phys. Rev. E} {\bf 2006}, {\em 74}, 16305--16311.
\bibitem{hahm} {Hahm, T.S.} {Nonlinear gyrokinetic equations for tokamak microturbulence.} {\em Phys. Fluids} {\bf 1988}, {\em 31}, 2670--2673.
\bibitem{Zweben} {Zweben, S.J.} {Search for coherent structure within tokamak plasma turbulence.} {\em Phys. Fluids} {\bf 2007}, {\em 28}, 974--982.
\bibitem{Naulin} {Naulin V.} {Turbulent transport and the plasma edge.} {\em J. Nucl. Mater.} {\bf 2007}, {\em 363-365}, 24-31.
\bibitem{kaye} {Kaye, S.M.; Barnes, C.W.; Bell, M.G.; DeBoo, J.C.; Greenwald, M.; Riedel, K.; Sigmar, D.; Uckan, N.; Waltz, N.} {Status of global energy confinement studies.} {\em Phys. Plasmas} {\bf 1990}, {\bf 2}, 2926--2940.
\bibitem{Lopez} {Cardozo, N.J.L.} {Perturbative transport studies in fusion plasmas.} {\em Plasma Phys. Control. Fusion} {\bf 1995}, {\em 37}, 799--852.
\bibitem{Gentle} {Gentle, K.W.;  Bravenec, R.V.;  Cima, G.; Gasquet, H.; Hallock, G.A.; Phillips, P.E.; Ross, D.W.; Rowan, W.L.; Wootton, A.J.} {An experimental counter‐example to the local transport paradigm.} {\em Phys. Plasmas} {\bf 1995}, {\em 2}, 2292--2298.
\bibitem{Mantica} {Mantica, P.; Galli, P.; Gorini, G.; Hogeweij, G.M.D.; de Kloe, J.; Cardozo, N.J.L.; RTP Team.} {Nonlocal transient transport and thermal barriers in rijnhuizen tokamak project plasmas.} {\em Phys. Rev. Lett.} {\bf 1999}, {\em 82}, 5048--5051.
\bibitem{van-Milligen} {Van-Milligen, B.P.; de la Luna, E.; Tabars, F.L.; Ascasíbar, E.; Estrada, T.; Castejón, F.; Castellano, J.; Cortés,~I.G.; Herranz, J.; Hidalgo, C.; et al.} {Ballistic transport phenomena in TJ-II.} {\em Nucl. Fusion} {\bf 2002}, {\em 42}, 787--795.
\bibitem{anderson2017} {Anderson, J.; Hnat, B.} {Statistical analysis of Hasegawa-Wakatani turbulence.} {\em Phys. Plasmas} {\bf 2017}, {\em 24}, 62301--62308.
\bibitem{moradi2018} {Moradi, S.; Anderson, J.; Romanelli, M.} {Evidence of non-local heat transport model in JET plasmas. Presented at EU-US Transport Task Force Meeting, Seville, Spain, 11--14 September 2018.}
\bibitem{schlesinger1993} {Schlesinger, M.F.; Zaslavsky, G.M.; Klafter, J.} {Strange kinetics.} {\em Nature} {\bf 1993}, {\em 363}, 31--37.
\bibitem{sokolov2002} {Sokolov, I.M.; Klafter, J.; Blumen, A.} {Fractional kinetics.} {\em Phys. Today} {\bf 2002}, {\em 55}, 48--54.
\bibitem{klafter2005} {Klafter, J.; Sokolov, I.M.} {Anomalous diffusion spreads its wings.} {\em Phys. World} {\bf 2005} {\em 18}, 29--32.
\bibitem{metzler2000} {Metzler, R.; Klafter, J.} {The~random walk's guide to anomalous diffusion: A fractional dynamics approach.} {\em Phys. Rep.} {\bf 2000}, {\em 339}, 1--77.
\bibitem{metzler2004} {Metzler, R.;  Klafter, J.} {The~restaurant at the end of the random walk: recent developments in the description of anomalous transport by fractional dynamics.} {\em J. Phys. A} {\bf 2004}, {\em 37}, 161--208.
\bibitem{mandelbrot1982} {Mandelbrot, B.B.} {\em Fractals and Geometry of Nature;} W. H. Freeman and Company: San Francisco, CA, USA, 1982; pp. 170--180.
\bibitem{anderson2010} {Anderson, J.; Xanthopoulos, P.} {Signature of a universal statistical description for drift-wave plasma turbulence.} {\em Phys. Plasmas} {\bf 2010}, {\em 17}, 110702--110704.
\bibitem{kim2009} {Kim, E.; Liu, H.L.; Anderson, J.} {Probability distribution function for self-organization of shear flows.} {\em Phys.~Plasmas} {\bf 2009} {\em 16}, 52301--52304.
\bibitem{moradipop2011} {Moradi, S.; Anderson J.; Weyssow, B.} {A theory of non-local linear drift wave transport.} {\em Phys. Plasmas} {\bf 2011}, {\em 18}, 062101--062106.
\bibitem{moradipop2012} {Moradi, S.; Anderson, J.} {Non-local gyrokinetic model of linear ion-temperature-gradient modes.} {\em Phys.~Plasmas} {\bf 2012}, {\em 19}, 82301--82307.
\bibitem{anderson2014} {Anderson, J.; Kim, E.; Moradi, S.} {A fractional Fokker--Planck model for anomalous diffusion.} {\em Phys. Plasmas} {\bf 2014}, {\em 21}, 122101--122108.
\bibitem{moradipop2016} {Moradi, S.; del Castillo, N.D.; Anderson, J.} {Charged particle dynamics in the presence of non-Gaussian Lévy electrostatic fluctuations.} {\em Phys. Plasmas} {\bf 2016}, {\em 23}, 907041--907045.
\bibitem{montroll1973} {Montroll, E.W.; Scher, H.} {Random walks on lattices. IV. Continuous-time walks and influence of absorbing boundaries.} {\em J. Stat. Phys.} {\bf 1973}, {\em 9}, 101--135.
\bibitem{kou2004} {Kou, S.C.; Sunney X.} {Generalized langevin equation with fractional Gaussian noise: Subdiffusion within a single protein molecule.} {\em Phys. Rev. Lett.} {\bf 2004}, {\em 93}, 1806031--1806034.
\bibitem{zaslavsky} {Combescure, M.} {\em Hamiltonian Chaos and Fractional Dynamics;} {Oxford University Press: Oxford, UK, 2005;} {pp.~0305--4470.}
\bibitem{del-Castillo-Negrete2004} {del Castillo, N.D.; Carreras, B.A.; Lynch, V.E.} {Fractional diffusion in plasma turbulence.} {\em Phys. Plasmas} {\bf 2004}, {\em 11}, 3854--3864.
\bibitem{del-Castillo-Negrete2010} {Del Castillo, N.D.} {Non-diffusive, non-local transport in fluids and plasmas.} {\em Nonlinear Proc. Geophys.} {\bf 2010}, {\em 17}, 795--807.
\bibitem{zaslavsky2002} {Zaslavsky, G.M.} {Chaos, fractional~kinetics, and~anomalous transport.} {\em Phys. Rep.} {\bf 2002}, {\em 371}, 461--580.
\bibitem{tarasov2005} {Tarasov, V.E.} {Fractional Liouville and BBGKI equations.} {\em J. Phys.} {\bf 2005}, {\em 7}, 17--33.
\bibitem{tarasov2006} {Tarasov, V.E.} {Fractional statistical mechanics.} {\em Chaos} {\bf 2006}, {\em 16}, 331081--331087.
\bibitem{levy}  {L\'{e}vy, P.} {{\em Théorie de L'addition des Variables Aléatoires;}} {Gauthier-Villiers: Paris, France, 1937.}
\bibitem{seshadri} {West, B.J.; Seshadri, V.} {Linear systems with Lévy fluctuations.} {\em Physical A} {\bf 1982}, {\em 113,} 203--216.
\bibitem{fogedby1994} {Fogedby, H.C.} {Langevin equations for continuous time Lévy flights.} {\em Phys. Rev. E} {\bf 1994}, {\em 50}, 1657--1660.
\bibitem{fogedby19942} {Fogedby, H.C.} {Lévy Flights in Random Environments.} {\em Phys. Rev. Lett.} {\bf 1994}, {\em 73}, 2517--2520.
\bibitem{diethelm} {Diethelm, K.; Freed, A.D. } {The~Fractional PECE Subroutine for the numerical solution of differential equations of fractional order}. In \emph{Forschung und Wissenschaftliches Rechnen}; Heinzel, S., Plesser, T., Eds.; Gessellschaft fur Wissenschaftliche Datenverarbeitung: Gottingen, Germany,  1999; pp. 57--71.
\bibitem{tsallis1995lnp} {Tsallis, C.; de Souza, A.M.C.; Maynard, R.} {Derivation of Lévy-type anomalous superdiffusion from generalized statistical mechanics.} {In~\em Lévy Flights and Related Topics in Physics;} {Springer: New York, NY,  USA, 1995;} {Volume 450,} \linebreak{pp. 269--289}.
\bibitem{tsallis1996} {Tsallis, C.; Bukman, D.J.} {Anomalous diffusion in the presence of external forces: Exact time-dependent solutions and their thermostatistical basis.} {\em Phys. Rev. E} {\bf 1996}, {\em 54}, 2197--2200.
\bibitem{tsallis1998} {Tsallis, C.; Mendes, R.S.; Plastino, A.R.} {The~role of constraints within generalized nonextensive statistics.} {\em Physical A} {1998}, {\bf 261}, 534--554.
\bibitem{hamza} {Hamza, A.B.; Krim, H.} {Jensen-R\'enyi divergence measure: Theoretical and Computational Perspectives. In~Proceeding of the 2003 IEEE International Symposium on Information Theory, Yokohama, Japan, 29~June--4~July~2003.} 
\bibitem{barkai} {Barkai, E.} {{Stable equilibrium based on Lévy statistics: Stochastic collision models approach.}} {\em {Phys. Rev. E.}} {\bf 2003}, {\em 68}, 551041--551044.
\bibitem{angulo} {Angulo, J.M.; Esquivel, F.J.} {Multifractal dimensional dependence assessment based on Tsallis mutual information.} {\em Entropy} {\bf 2015}, {\em 17}, 5382--5401.
\bibitem{balasis2011} {Balasis, G.; Daglis, I.A.; Anastasiadis, A.; Papadimitriou, C.; Mandea, M.; Eftaxiasb, K.} {Universality in solar flare, magnetic~storm and earthquake dynamics using Tsallis statistical mechanics.} {\em Physical A} {\bf2011}, {\em 390}, 341--346.
\bibitem{pavlos2012} {Pavlos, G.P.; Karkatsanis, L.P.; Xenakis, M.N.; Sarafopoulos D.; Pavlos, E.G.} {Tsallis statistics and magnetospheric self-organization.} {\em Physical A} {\bf 2012}, {\em 391}, 3069--3080.
\bibitem{pavlos20122} {Pavlos, G.P.; Karkatsanis, L.P.; Xenakis, M.N.} {Tsallis non-extensive statistics, intermittent~turbulence, SOC~and chaos in the solar plasma, Part one: Sunspot dynamics.} {\em Physical A} {\bf 2012}, {\em 391}, 6287--6319.
\bibitem{tsallis1995} {Tsallis, C.; L\'{e}vy, S.V.F.; Souza, A.M.C.; Maynard, R.} {Statistical-mechanical foundation of the ubiquity of Lévy distributions in nature.} {\em Phys. Rev. Lett.} {\bf 1995}, {\em 75}, 3589--3593.
\bibitem{prato1999} {Prato, D.; Tsallis, C.} {Nonextensive foundation of Lévy distributions.} {\em Phys. Rev. E} {\bf 1999}, {\em 60}, 2398--2401.
\bibitem{milovanov} {Milovanov, A.V.; Rasmussen, J.J.} {Lévy flights on a comb and the plasma staircase.} {\em Phys. Rev. E} {\bf 2018}, {\em 98}, 222081--222092.
\bibitem{jespersen} {Jespersen, S.; Metzler, R.; Fogedby, H.C.} {Lévy flights in external force fields: Langevin and fractional Fokker--Planck equations and their solutions.} {\em Phys. Rev. E} {\bf 1999}, {\em 59}, 2736--2745.
\end{thebibliography}
\end{document}